\documentclass[a4paper,10pt]{article}

\usepackage{graphicx,amsfonts,amssymb}

\begin{document}

\begin{center}



{\Large\textbf{Two component integrable systems modelling shallow
water waves: the constant vorticity case
\\}} \vspace {10mm} \vspace{1mm} \noindent

{\large \bf Rossen I. Ivanov}\footnote{E-mail: rivanov@dit.ie}

\hskip-.8cm

\hskip-1cm {\it School of Mathematical Sciences,
Dublin Institute of Technology, \\
Kevin Street, Dublin 8, Ireland }\\
\hskip-.8cm

\hskip-.8cm

\begin{abstract}
In this contribution we describe the role of several two-component
integrable systems in the classical problem of shallow water
waves. The starting point in our derivation is the Euler equation
for an incompressible fluid, the equation of mass conservation,
the simplest bottom and surface conditions and the constant
vorticity condition. The approximate model equations are generated
by introduction of suitable scalings and by truncating asymptotic
expansions of the quantities to appropriate order. The so obtained
equations can be related to three different integrable systems: a
two component generalization of the Camassa-Holm equation, the
Zakharov-Ito system and the Kaup-Boussinesq system.

The significance of the results is the inclusion of vorticity, an
important feature of water waves that has been given increasing
attention during the last decade. The presented investigation
shows how -- up to a certain order -- the model equations relate
to the shear flow upon which the wave resides. In particular, it
shows exactly how the constant vorticity affects the equations.

\end{abstract}


\hskip-.8cm


\end{center}


\section{Introduction}
\label{s1}

The integrable nonlinear equations are used extensively as
approximate models in hydrodynamics. They describe in a relatively
simple way the competition between nonlinear and dispersive
effects. The best known example in this regard is the Korteweg-de
Vries (KdV) equation. The use of term { \it integrable}
corresponds to the idea that such equations are in some sense
exactly solvable and exhibit global regular solutions. This
feature is very important for applications where in general
analytical results are preferable to numerical computations.

The Camassa-Holm (CH) and Degasperis-Procesi (DP) equations
\cite{CH93,DP,DHH} are another two integrable equations with
application in the theory of water waves \cite{CH93,
DGH03,J02,J03,J03a,CL09,I07}. The excitement that greeted the CH
and DP equations is due to their non-standard properties that set
them apart from the classical soliton equations such as KdV. The
first most remarkable of these properties is the presence of
multi-soliton solutions consisting of a train of peaked solitary
waves (or 'peakons') \cite{CH93,CS00}. Another remarkable property
of the CH and DP equations is the occurrence of breaking waves
\cite{CH93,CE,McK04,ELY07a,C00} (i.e. a solution that remains
bounded while its slope becomes unbounded in finite time \cite{W})
as well as that of smooth solutions defined for all times
\cite{CS00,C,David08}.

In many recent publications the problem of water waves with
nonzero vorticity and especially with constant vorticity is under
investigation - e.g. see the publications
\cite{J03,C05,CSS06,CJ08a,CIP08,CEW07,W07,GW08,ME08,H08,V-B95} and
the references therein. The nonzero vorticity case arises for
example in situations with underlying shear flow \cite{J03}.

Our aim is to describe the derivation of shallow water model
equations for the constant vorticity case and to demonstrate how
these equations can be related to some other integrable systems: a
two component generalization of the Camassa-Holm equation
\cite{OR}, Zakharov-Ito system \cite{Z,Ito} and Kaup-Boussinesq
system \cite{K76}. A starting point in our derivation are the
equations that express the constant vorticity and the mass
conservation. Another approach, based on the Green-Naghdi
approximation is used in an alternative derivation of the two
component Camassa-Holm equation in \cite{CI08}, where the
occurrence of solutions in the form of breaking waves is also
established. The Kaup-Boussinesq system is used as a model in
hydrodynamics under slightly different assumptions in
\cite{W,K76,EGP01}.

\section{Governing equations for the inviscid fluid motion}
\label{s2}

The motion of inviscid fluid with a constant density $\rho$ is
described by the Euler's equations:

\begin{eqnarray} \frac{\partial {\bf v}}{\partial t}+({\bf v}\cdot\nabla){\bf
v}&=&-\frac{1}{\rho}\nabla P+ {\bf g}, \nonumber \\
\nabla \cdot {\bf v}&=& 0, \nonumber
\end{eqnarray} where ${\bf v}(x,y,z,t)$ is the velocity of the
fluid at the point $(x,y,z)$ at the time $t$, $P(x,y,z,t)$ is the
pressure in the fluid, ${\bf g}=(0,0,-g)$ is the constant Earth's
gravity acceleration.

Consider now a motion of a shallow water over a flat bottom, which
is located at $z=0$ (Fig. 1). We assume that the motion is in the
$x$-direction, and that the physical variables do not depend on
$y$.

\begin{figure}[htp]
\centering
\includegraphics[width=0.7\textwidth]{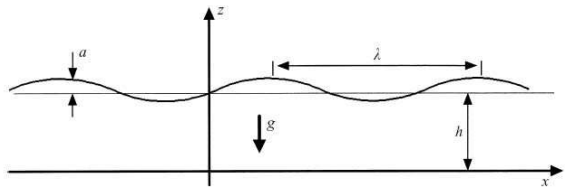}
\caption{Water waves: general notations}\label{fig:1}
\end{figure}


Let $h$ be the mean level of the water and let $\eta (x,t)$
describes the shape of the water surface, i.e. the deviation from
the average level. The pressure is $ P(x,z,t)=P_A+\rho g
(h-z)+p(x,z,t)$, where $P_A$ is the constant atmospheric pressure,
and $p$ is a pressure variable, measuring the deviation from the
hydrostatic pressure distribution.

On the surface $z=h+\eta$, $P=P_A$ and therefore $p=\eta \rho g$.
Taking ${\bf v}\equiv(u,0,w)$ we can write the kinematic condition
on the surface as $ w= \eta_t+ u \eta_x$ on $ z=h+\eta$
\cite{J97}. Finally, there is no vertical velocity at the bottom,
thus $ w=0 $ on $ z=0 $. All these equations can be
written as a system  \begin{eqnarray}   u_t&+&uu_x+wu_z=-\frac{1}{\rho}p_x, \nonumber \\
 w_t&+&uw_x+ww_z=-\frac{1}{\rho}p_z, \nonumber \\
 u_x&+&w_z = 0, \nonumber \\
w&=& \eta_t+ u \eta_x, \quad  p=\eta \rho g \quad {\rm on}\quad
z=h+\eta, \nonumber \\
w &=& 0 \quad {\rm on} \quad z=0. \nonumber
\end{eqnarray}

Let us introduce now dimensionless parameters $\varepsilon=a/h$
and $\delta=h/\lambda$, where $a$ is the typical amplitude of the
wave and $\lambda$ is the typical wavelength of the wave. Now we
can introduce dimensionless quantities, according to the magnitude
of the physical quantities, see \cite{J02,J97,CJ08} for details: $
x\rightarrow\lambda x$,   $z\rightarrow z h$, $t\rightarrow
\frac{\lambda}{\sqrt{g h}}t$, $ \eta\rightarrow a\eta$,
$u\rightarrow \varepsilon \sqrt{g h} u$,  $ w\rightarrow
\varepsilon \delta \sqrt{g h} w$, $p\rightarrow \varepsilon \rho g
h $. This scaling is due to the observation that both $w$ and $p$
are proportional to $\varepsilon$ i.e. the wave amplitude, since
at undisturbed surface ($\varepsilon=0$) both $w=0$ and $p=0$. The
system in the new, dimensionless variables is

\begin{eqnarray} u_t&+&\varepsilon (uu_x+wu_z)=-p_x,  \nonumber \\
\delta^2&(&w_t+\varepsilon (uw_x+ww_z))=-p_z,  \nonumber \\
u_x&+&w_z=0,  \nonumber \\
w&= &\eta_t+ \varepsilon u \eta_x,\quad  p=\eta \quad
{\rm on}\quad z=1+\varepsilon \eta, \nonumber \\
w&=&0 \quad {\rm on} \quad z=0. \nonumber \end{eqnarray}

\section{ Waves in the presence of shear}

So far no assumptions have been made on the presence of shear. Now
let us notice that there is an exact solution of the governing
equations of the form $u=\tilde{U}(z)$, $0\leq z\leq h$, $w\equiv
0$, $p\equiv 0$, $\eta \equiv 0$. This solution represents an
arbitrary underlying 'shear' flow \cite{J03}. Let us consider
waves in the presence of a shear flow. In such case the horizontal
velocity of the flow will be $\tilde{U}(z)+ u $.  The scaling for
such solution is clearly $ u\rightarrow \sqrt{g
h}\Big(\tilde{U}(z)+\varepsilon u\Big)$, where $u$ on the
left-hand side is the horizontal velocity before the initial
scaling, and the scaling for the other variables is as before.
Thus, in this case we have (the prime denotes derivative with
respect to $z$): \begin{eqnarray}
u_t&+&\tilde{U}u_x+w\tilde{U}'+\varepsilon
(uu_x+wu_z)= -p_x, \label{E1} \\
 \delta^2&(&w_t+\tilde{U}w_x+\varepsilon
(uw_x+ww_z))= -p_z,\label{E2}\\
u_x&+&w_z=0, \label{E3}\\
w&=& \eta_t+ (\tilde{U}+\varepsilon u )\eta_x,  \quad p=\eta \quad
{\rm on} \quad z=1+\varepsilon \eta \label{E4} \\
w&=&0 \quad {\rm on}\quad z=0.\label{E5} \end{eqnarray}

We consider only the simplest nontrivial case: a linear shear,
$\tilde{U}(z)=Az$, where $A$ is a constant ( $0\le z\le 1$). We
choose $A>0$, so that the underlying flow is propagating in the
positive direction of the $x$-coordinate. Burns condition
\cite{B53} gives the following expression for $c$, the speed of
the travelling waves in linear approximation :
\begin{equation} c=\frac{1}{2}\Big(A\pm\sqrt{4+A^2}\Big).
\label{c(A)}\end{equation} This expression will be derived again
in our further considerations, e.g. see (\ref{c-A equation}). Note
that if there is no shear ($A=0$), then $c=\pm 1$.

Before the scaling the vorticity is $ \omega=(U+u)_z-w_x $ or in
terms of the rescaled variables ($\omega \rightarrow \sqrt{h/g}
\phantom{*} \omega $),
\begin{eqnarray} \omega=A + \varepsilon(u_z-\delta^2w_x).\nonumber
\end{eqnarray} We are looking for a solution with constant
vorticity $\omega=A$, and therefore we require that
\begin{equation} u_z-\delta^2w_x=0. \label{vortCond}\end{equation}

This assumption amounts to considering approximate wave-solutions
that are interactions of an underlying shear flow and an
irrotational disturbance thereof. From (\ref{vortCond}),
(\ref{E3}) and (\ref{E5}) we obtain
\begin{eqnarray}
u&=&u_0-\delta^2\frac{z^2}{2}u_{0xx}+\mathcal{O}(\varepsilon^2,
\delta^4, \varepsilon \delta^2), \label{u}\\
w&=&-zu_{0x}+\delta^2\frac{z^3}{6}u_{0xxx}+\mathcal{O}(\varepsilon^2,
\delta^4, \varepsilon \delta^2),\label{w} \end{eqnarray} where
$u_0(x,t)$ is the leading order approximation for $u$. Note that
$u_0$ does not depend on $z$ since from (\ref{vortCond}) it
follows that $u_{z}=0$ when $\delta\rightarrow 0$.

From (\ref{E4}) with (\ref{u}) and (\ref{w})  we obtain
\begin{equation} \eta_t+A\eta_x+\Big[(1+\varepsilon
\eta)u_0+\varepsilon
\frac{A}{2}\eta^2\Big]_x-\delta^2\frac{1}{6}u_{0xxx}=0,\label{Eta_t}\end{equation}
ignoring terms of order $\mathcal{O}(\varepsilon^2, \delta^4,
\varepsilon \delta^2)$.

From (\ref{E2}), (\ref{E4}), (\ref{u}) and (\ref{w})  we have
(again, ignoring terms of order $\mathcal{O}(\varepsilon^2,
\delta^4, \varepsilon \delta^2)$) \begin{eqnarray}
p=\eta-\delta^2\Big[ \frac{1-z^2}{2}u_{0xt}+\frac{1-z^3}{3}A
u_{0xx}\Big]. \nonumber \end{eqnarray}

Then (\ref{E1}) gives (note that there is no $z$-dependence!)
\begin{equation}  \Big(
u_0-\delta^2\frac{1}{2}u_{0xx}\Big)_t+\varepsilon u_0
u_{0x}+\eta_x -\delta^2 \frac{A}{3} u_{0xxx}=0.
\label{u0_t}\end{equation}  Letting both the parameters
$\varepsilon$ and  $\delta$ to $0$, we obtain from (\ref{Eta_t}),
(\ref{u0_t}) the system of linear equations \begin{eqnarray} u_{0t}&+&\eta_x=0, \label{1lin Eq} \\
\eta_t&+&A\eta_x+u_{0x}=0, \label{2lin Eq} \end{eqnarray} giving
\begin{eqnarray} \eta_{tt}+A\eta_{tx}-\eta_{xx}=0. \label{linEq}
\end{eqnarray} The linear equation (\ref{linEq}) has a travelling
wave solution $\eta=\eta(x-ct)$ with a velocity $c$ satisfying
\begin{equation} c^2-Ac-1=0. \label{c-A equation}
\end{equation} This gives the same solution for $c$ that follows
from the Burns condition (\ref{c(A)}). There is one positive and
one negative solution, representing left and right running waves.
We assume that we have only one of these waves, then (e.g. from
(\ref{1lin Eq}))
\begin{equation} \eta=cu_0+\mathcal{O}(\varepsilon, \delta^2).
\label{Eta/u Leadord}\end{equation}

Let us introduce a new variable\begin{equation} \rho=1+
\varepsilon \alpha \eta+\varepsilon^2 \beta \eta^2+\varepsilon
\delta^2 \gamma u_{0xx},\label{Rho}\end{equation} for some
constants $\alpha$, $\beta$ and $\gamma$. These constants will be
determined in our further considerations.  The variable $\rho$
will be used instead of $\eta$ as a tool for mathematical
simplification of our equations. The expansion of $\rho^2$ in the
same order of $\varepsilon$ and $\delta^2$ is
\begin{equation} \rho^2=1+\varepsilon
(2\alpha)\eta+\varepsilon^2(\alpha^2+2\beta) \eta^2+\varepsilon
\delta^2 (2\gamma) u_{0xx}. \label{Rho^2}\end{equation}

With this definition one can express $\eta$ in terms of $\rho $
and write equation (\ref{Eta_t}) in the form (keeping only terms
of order $\mathcal{O}(\varepsilon,\delta^2)$):
\begin{eqnarray}
\frac{\rho_{t}+A \rho_x }{\alpha \varepsilon}&+&
\delta^2\Big(\frac{\gamma}{\alpha}(c-A)-\frac{1}{6} \Big)u_{0xxx}
\nonumber \\
&+&\Big[ (1+ \varepsilon \eta)u_0 + \varepsilon \frac{A}{2} \eta^2
+ \varepsilon \frac{\beta}{\alpha} c u_0^2 \Big]_x=0. \label{Rho_t
1} \end{eqnarray} One can eliminate the $u_{0xxx}$-term by
choosing \begin{eqnarray} \frac{\gamma}{\alpha}=\frac{1}{6(c-A)}.
\label{gamma/alpha} \end{eqnarray}

\noindent Equation (\ref{Rho_t 1}) becomes \begin{eqnarray}
\frac{\rho_{t}+A \rho_x }{\alpha \varepsilon}+\Big[ \Big(1+
\varepsilon(1+\frac{Ac}{2}+\frac{\beta}{\alpha})\eta \Big)u_0
\Big]_x=0. \label{Rho_t 2} \end{eqnarray} With the choice
\begin{eqnarray} \alpha = 1+\frac{Ac}{2}+\frac{\beta}{\alpha}
\label{AlphaCond} \end{eqnarray} we can write (\ref{Rho_t 2}) in
the form \begin{eqnarray} \rho_{t}+A \rho_x +\alpha
\varepsilon(\rho u_0)_x=0, \label{Rho_t} \end{eqnarray} which
contains only the variables $\rho$ and $u_0$ but not $\eta$.

\section{Two component Camassa-Holm system }

In this section we proceed with our derivation in a direction that
leads to a two component Camassa-Holm system. Expressing $\eta$ in
terms of $\rho $ in (\ref{u0_t}) we obtain (matching terms of
order $\mathcal{O}(\varepsilon,\delta^2)$):

\begin{eqnarray}
m_{t}+Am_x-Au_{0x}&+&\delta^2\Big(\frac{A}{6}+\kappa(A-c)-\frac{\gamma}{\alpha}\Big)u_{0xxx}
\nonumber \\
&+&\varepsilon \Big(1-\frac{\alpha^2+2\beta}{\alpha}c^2 \Big)u_0
u_{0x}+\frac{1}{2\varepsilon \alpha}(\rho^2)_x=0,
\label{m_1}\end{eqnarray}

\noindent where  $m=u_0-\delta^2(\frac{1}{2}+\kappa) u_{0xx}$,
$\kappa$ is arbitrary: we are adding and subtracting $\delta^2
\kappa u_{0xxt}$, making use of (\ref{Eta/u Leadord}). Fixing
\begin{equation} \kappa=\frac{1}{A-c}\Big(
\frac{\gamma}{\alpha}-\frac{A}{6}\Big) \label{kapa1}
\end{equation} leads to the disappearance of the  $u_{0xxx}$ - term .

The relations (\ref{gamma/alpha}) and (\ref{kapa1}) give
\begin{eqnarray} \kappa=\frac{1}{6(c-A)}\Big(A-\frac{1}{c-A}\Big).
 \label{Kappa}\end{eqnarray}

Thus equation (\ref{m_1}) can be written as (matching only terms
up to the order $\mathcal{O}(\varepsilon,\delta^2)$)
\begin{eqnarray} m_{t}+Am_x-Au_{0x}+\varepsilon \frac{1}{3}
\Big(1-\frac{\alpha^2+2\beta}{\alpha}c^2 \Big)[2m u_{0x}+ u_0
m_{x}]+\frac{\rho \rho_x}{\varepsilon \alpha}=0. \label{m_t}
\end{eqnarray}

Recall that $m=u_0-\delta^2Bu_{0xx}$, where (see (\ref{Kappa}) and
(\ref{c(A)}))
\begin{eqnarray}
B=\kappa+\frac{1}{2}=\frac{A^2-c^2+2}{3(c-A)^2}=\frac{1}{3c^2(c-A)^2}.
\label{B} \end{eqnarray} Note that $B$ is always positive and the
denominator in (\ref{B}) nonzero  (since $c \ne A$ -- see
(\ref{c-A equation})).  The rescaling $u_0\rightarrow
\frac{1}{\alpha \varepsilon} u_0$, $x\rightarrow
\frac{\delta}{\sqrt{B}}\phantom{*} x$, $t\rightarrow
\frac{\delta}{\sqrt{B}} t$ in (\ref{m_t})  and (\ref{Rho_t}) is
now only for the sake of mathematical clarity and simplicity and
gives: \begin{eqnarray} m_{t}&+&Am_x-Au_{0x}+\frac{1}{3\alpha}
\Big(1-\frac{\alpha^2+2\beta}{\alpha}c^2 \Big)[2m u_{0x}+ u_0
m_{x}]+\rho \rho_x=0, \nonumber \\
m&=&u_0-u_{0xx},  \nonumber \\
\rho_{t}&+&A \rho_x +(\rho u_0)_x=0. \nonumber \end{eqnarray}

Finally, we choose \begin{eqnarray} \frac{1}{3\alpha}
\Big(1-\frac{\alpha^2+2\beta}{\alpha}c^2 \Big)=1 \label{constr
alpha beta gamma}\end{eqnarray} and thus \begin{eqnarray}
m_{t}&+&Am_x-Au_{0x}+2m u_{0x}+ u_0
m_{x}+\rho \rho_x=0, \quad m=u_0-u_{0xx}, \\
\rho_{t}&+&A \rho_x +(\rho u_0)_x=0. \end{eqnarray}

The constants $\alpha$, $\beta$ and $\gamma$ can be determined
from the constraints (\ref{AlphaCond}), (\ref{constr alpha beta
gamma}) and (\ref{gamma/alpha}): \begin{eqnarray}
\alpha&=&\frac{1}{3(1+c^2)}+\frac{c^2}{3}, \label{alpha} \\
\beta&=&\Big[\frac{1}{3(1+c^2)}-\frac{3+c^2}{6}\Big]\alpha , \label{beta}\\
\gamma &=& \frac{1}{6(c-A)}\alpha. \label{gamma} \end{eqnarray}

\noindent Note that from (\ref{alpha}) it follows that $\alpha$ is
always positive.

Now let us express the original variable $\eta $ in terms of the
'auxiliary' variable $\rho$. Before the rescaling we had $\alpha
\varepsilon \eta = \rho-1-\varepsilon ^2 \beta c^2 u_0^2 -
\varepsilon\delta^2 \gamma u_{0xx}$. Since in the leading order
$\eta=cu_0$ the rescaling of $\eta$ is $\eta \rightarrow
\frac{1}{\alpha \varepsilon} \eta$, thus in terms of the rescaled
variables

\begin{eqnarray} \eta = \rho-1- \frac{\beta c^2}{\alpha^2}   u_0^2 -
B \frac{\gamma}{\alpha}u_{0xx}. \nonumber
\end{eqnarray}

With a Galilean transformation (that we use only to simplify our
equations and to bring them to the form that is widely used), such
that $\partial_{t'}=\partial_t+A\partial_x$,
$\partial_{x'}=\partial_x$ ($x'= x-At$, $t'=t$) we obtain
\begin{eqnarray} m_{t'}& -&Au_{0x'}+2m u_{0x'}+ u_0 m_{x'}+\rho
\rho_{x'}=0,  \quad
m=u_0-u_{0x'x'} \label{3'}\\
\rho_{t'} &+&(\rho u_0)_{x'}=0. \label{4'} \end{eqnarray}

The system (\ref{3'}), (\ref{4'}) is an integrable 2-component
Camassa-Holm system that appears in \cite{OR}, generalizing the
famous Camassa-Holm equation \cite{CH93}.

Let us drop the primes and write it in the form
\begin{eqnarray} m_{t}& -&Au_{0x}+2m u_{0x}+ u_0 m_{x}+\rho \rho_{x}=0,  \quad
m=u_0-u_{0xx} \label{3}\\
\rho_{t} &+&(\rho u_0)_{x}=0. \label{4} \end{eqnarray} It
generalizes the Camassa-Holm equation \cite{CH93} in a sense that
it can be obtained from it via the obvious reduction $\rho\equiv
0$. The system is integrable, since it can be written as a
compatibility condition of two linear systems (Lax pair) with a
spectral parameter $\zeta$:
\begin{eqnarray} \Psi_{xx}=\Big(-\zeta^2\rho^2+\zeta
(m-\frac{A}{2}) +\frac{1}{4}\Big)\Psi, \qquad
\Psi_{t}=\Big(\frac{1}{2\zeta}-u_0\Big)\Psi_x+\frac{1}{2}u_{0x}\Psi.
\nonumber
\end{eqnarray} The system is also bi-Hamiltonian. The first Poisson bracket is
\begin{eqnarray} \{F_1,F_2\}=-\int\Big[\frac{\delta F_1}{\delta
m}(-A\partial+m\partial+\partial m)\frac{\delta F_2}{\delta
m}+\frac{\delta F_1}{\delta m}\rho
\partial\frac{\delta F_2}{\delta \rho}+\frac{\delta F_1}{\delta \rho}\partial \rho \frac{\delta F_2}{\delta
m}\Big]{\rm d}x \nonumber \end{eqnarray} for the Hamiltonian
$H_1=\frac{1}{2}\int (u_0(m-\frac{A}{2})+\rho^2){\rm d}x$.

The second Poisson bracket is \begin{eqnarray}
\{F_1,F_2\}_2=-\int\Big[\frac{\delta F_1}{\delta
m}(\partial-\partial^{3})\frac{\delta F_2}{\delta m}+\frac{\delta
F_1}{\delta \rho}
\partial\frac{\delta F_2}{\delta \rho}\Big]{\rm d}x \nonumber \end{eqnarray} for the Hamiltonian $H_2=\frac{1}{2}\int
(u_0\rho^2+u_0^3+u_0u_{0x}^2-Au_0^2){\rm d}x. $ It has two
Casimirs: $\int \rho {\rm d}x$ and $\int m {\rm d} x$.

The system has an interesting interpretation in group-theoretical
context. The first Poisson bracket gives rise to a Lie-algebraic
structure. This fact is well studied in the case $\rho\equiv 0$
when the system coinsides with the Camassa-Holm equation
\cite{M98,HMR98,CK03,CK06,K04}. Then the corresponding Lie algebra
is the Virasoro algebra.

By considering the expansions
\begin{eqnarray}m=\frac{1}{2\pi}\sum_{n \in \mathbb{Z}}L_ne^{inx},
\qquad \rho=\frac{1}{2\pi}\sum_{n \in
\mathbb{Z}}\rho_ne^{inx}\nonumber \end{eqnarray} we obtain the
following Lie-algebra with respect of the first Poisson bracket:

\begin{eqnarray} i\{L_n,L_k\}&=&(n-k)L_{n+k}-2\pi A n \delta_{n+k},
\label{Vir-1} \\
i\{\rho_n,L_k\}&=& n\rho_{n+k}, \label{Vir-2} \\
i\{\rho_n,\rho_k\}&=& 0  \label{Abelian Alg}\end{eqnarray}

The Lie algebra (\ref{Vir-1}) -- (\ref{Abelian Alg}) is a
semidirect product of the Virasoro algebra ($\frak{vir}$)
(\ref{Vir-1}) with a central charge proportional to $A$, and the
abelian algebra $C^{\infty}(\mathbb{R})$ (\ref{Abelian Alg})
\cite{H88}. Note that the central extension of the Virasoro
algebra  contains only $An$ term but not $An^3$ term since the
Hamilton operator contains only the first derivative $A\partial$.
The system (\ref{3}), (\ref{4}) represents the equations of the
geodesic motion on the corresponding Lie group $\frak{Vir}\ltimes
C^{\infty}(\mathbb{R})$ for the metric, defined by $H_1$:
\begin{eqnarray} || (u_0, \rho)||^2= \int
(u_0^2+u_{0x}^2+\rho^2) {\rm d}x, \nonumber
\end{eqnarray} which is right-invariant under the natural group action.

\section{Zakharov-Ito system}

In this section we describe a derivation that matches the
approximate equations (\ref{u0_t}), (\ref{Rho_t}) (where $\rho$ is
given in (\ref{Rho})) to the integrable Zahkarov-Ito system
\cite{Z,Ito,Kup,Boiti}:

\begin{eqnarray} u_{0t}&-&4k u_{0x} +u_{0xxx} +3u_0u_{0x}+\rho
\rho_x=0.
\label{ZI1}\\
\rho_t&+&(u_0\rho)_x=0, \label{ZI2}
\end{eqnarray} where $k$ is an arbitrary constant. The system is
formally integrable by the virtue of the Lax pair \cite{Bog,TW05}

\begin{eqnarray} \Psi_{xx}=\Big(\zeta-\frac{u_0}{2}+k-\zeta^{-1}\frac{\rho^2}{16}\Big)\Psi,
\qquad \Psi_{t}=-(4\zeta+u_0)\Psi_x+\frac{1}{2}u_{0x}\Psi.
\nonumber
\end{eqnarray}

Equation (\ref{u0_t}) can be written in the form (cf. (\ref{Eta/u
Leadord}) and (\ref{1lin Eq}))

\begin{equation}
u_{0t}+\delta^2Ku_{0xxx}+\varepsilon
\Big(1-\frac{\alpha^2+2\beta}{\alpha}\Big)u_0 u_{0x}+\frac{\rho
\rho_x}{\alpha\varepsilon}=0. \label{u0_tZI1}\end{equation}

\noindent where $K$ is a constant, given by [see
(\ref{gamma/alpha}) and (\ref{c-A equation})]
\begin{eqnarray} K=
\frac{c}{2}-\frac{A}{3}-\frac{\gamma}{\alpha}=\frac{1}{3}(c-A).
\end{eqnarray}

For one of the roots of the equation in (\ref{c(A)}), $K$ is
positive and for the other it is negative. We can fix the
constants $\alpha$, $\beta$ and $\gamma$ by the conditions
(\ref{AlphaCond}) and
\begin{eqnarray} \frac{1}{3\alpha}
\Big(1-\frac{\alpha^2+2\beta}{\alpha}c^2 \Big)=1, \nonumber
\end{eqnarray} which are formally the same as those for the
Camassa-Holm system, giving the same expressions (\ref{alpha}) --
(\ref{gamma}).

The rescaling $u_0\rightarrow \frac{1}{\alpha \varepsilon} u_0$,
$x\rightarrow \frac{\delta}{\sqrt{K}} x$, $t\rightarrow
\frac{\delta}{\sqrt{K}} t$ in (\ref{u0_tZI1})  and (\ref{Rho_t})
gives: \begin{eqnarray} u_{0t}&+&Au_{0x}-Au_{0x}+u_{0xxx}+3u_0
u_{0x}+ \rho \rho_x=0, \nonumber \\
\rho_{t}&+&A \rho_x +(\rho u_0)_x=0. \nonumber \end{eqnarray}

\noindent Note that a coordinate change $(x,t)\rightarrow i(x,t)$
maps into another system with real variables and therefore if
$K<0$ we still can apply formally the above rescaling. One can use
a Galilean transformation $x'= x-At$, $t'=t$ to obtain

\begin{eqnarray} u_{0t'}&-&Au_{0x'}+u_{0x'x'x'}+3u_0
u_{0x'}+ \rho \rho_{x'}=0, \nonumber \\
\rho_{t'}&+&(\rho u_0)_{x'}=0, \nonumber \end{eqnarray}

that matches the Zakharov-Ito system (\ref{ZI1}), (\ref{ZI2}) if
the constant is chosen to be $k=A/4$. The 'physical' variable
$\eta $ in terms of the 'auxiliary' $\rho$ and $u_0$ (for the
rescaled variables) is \begin{eqnarray} \eta = \rho-1- \frac{\beta
c^2}{\alpha^2}   u_0^2 - K \frac{\gamma}{\alpha}u_{0xx}. \nonumber
\end{eqnarray}

\section{Kaup-Boussinesq system}

Another integrable system matching the water waves asymptotic
equations to the first order of the small parameters $\varepsilon,
\delta$ is the Kaup-Boussinesq system \cite{K76,EGP01}. In this
section we describe briefly its derivation. Introducing
\begin{eqnarray} V=u_0-\delta^2 \Big(\frac{1}{2}-\frac{A}{3c}\Big)
u_{0xx}\equiv u_0-\delta^2 \Big(\frac{1}{6}+\frac{1}{3c^2}\Big)
u_{0xx} \nonumber \end{eqnarray} the equation (\ref{u0_t}) can be
written as
\begin{equation} V_t+\varepsilon V V_x+ \eta_x=0 . \label{KB1} \end{equation}

Equation (\ref{Eta_t}) in the first order in $\varepsilon,
\delta^2$ is \begin{eqnarray} \eta_t+\Big[A\eta+(1+\varepsilon
\eta)u_0+\varepsilon
\frac{A}{2}\eta^2\Big]_x-\delta^2\frac{1}{6}u_{0xxx}=0 \nonumber
\end{eqnarray} and with a shift $\eta\rightarrow \eta - \frac{1}{\varepsilon}$
it becomes \begin{eqnarray} \eta_t+\varepsilon
(1+\frac{Ac}{2})(\eta u_0)_x -\delta^2\frac{1}{6}u_{0xxx}=0,
\nonumber \end{eqnarray}  or \begin{eqnarray} \eta_t+\varepsilon
\frac{1+c^2}{2}(\eta V)_x -\delta^2\frac{1}{6}V_{xxx}=0.
\label{KB2}
\end{eqnarray}

Further rescaling in (\ref{KB1}) and (\ref{KB2}) leads to  the
Kaup-Boussinesq system
\begin{eqnarray} V_t+ V V_x+ \eta_x=0, \qquad
\eta_t-\frac{1}{4}V_{xxx}+\frac{1+c^2}{2}(\eta V)_x=0, \nonumber
\end{eqnarray}

\noindent which is integrable iff $A=0$  ($c^2=1$) with a Lax pair
\begin{eqnarray}
\Psi_{xx}&=&-\Big((\zeta-\frac{1}{2}V)^2-\eta\Big)\Psi, \nonumber \\
\Psi_{t}&=&-(\zeta+\frac{1}{2}V)\Psi_x+\frac{1}{4}V_x\Psi.
\nonumber
\end{eqnarray}

The integrability of the system, as well as the Inverse Scattering
Method for it has been investigated firstly by D.J. Kaup
\cite{K76}. His motivation has been to derive an integrable
water-wave system with a second-order eigenvalue problem, which is
readily solvable in comparison to the third-order eigenvalue
problem for the Boussinesq equation. In our context, however, this
system is relevant only in the case with zero vorticity.

\section{Discussion}

Apparently the described method can be used for other
two-component integrable systems with a similar structure, e.g.
see the classification in \cite{TW05}. It is interesting to
investigate further which specific properties of the original
governing equations are preserved in the 'integrable' approximate
models. For example the 2-component Camassa-Holm system for
certain initial data admits wave breaking \cite{CI08}. Peakons do
not occur in the case $A=0$ \cite{CI08} and most certainly not in
the case $A\ne 0$, due to the term  with linear dispersion
$Au_{0x}$. However, in the 'short-wave limit' where $m=-u_{0xx}$
and $A=0$ peakon solutions are possible \cite{CI08}. Recently a
similar system with peakon solutions have been constructed in
\cite{HLT}.

The case with $-\rho\rho_x$ term (instead of $+\rho\rho_x$) in
(\ref{3}) is also integrable \cite{I06} and it is studied in
\cite{ELY07}. The two component Camassa-Holm system appears also
in plasma theory models \cite{GHT08,HT09} and in the theory of
metamorphosis \cite{HTY}. Other integrable multi-component
generalizations of the Camassa-Holm equation (including other
two-component ones) are constructed in \cite{I06}.



\section*{Acknowledgments}
The author is thankful to Prof. A. Constantin, Prof. R. Johnson
and Dr G. Grahovski for stimulating discussions and to both
referees for their comments and suggestions. Part of this work has
been done during the workshop 'Wave Motion' held in Oberwolfach,
Germany (8 -- 14 February 2009). Partial support from INTAS grant
No 05-1000008-7883 is acknowledged.

\end{document}